\documentclass[12pt]{article}
\usepackage{graphicx}
\begin{document}
\rightline{NKU-2013-SF2}
\bigskip

\newcommand{\be}{\begin{equation}}
\newcommand{\ee}{\end{equation}}
\newcommand{\noi}{\noindent}
\newcommand{\refb}[1]{(\ref{#1})}
\newcommand{\ra}{\rightarrow}

\begin{center}
{\Large\bf Born-Infeld-de Sitter gravity: Cold, ultracold and Nariai black holes}

\end{center}
\hspace{0.4cm}
\begin{center}
Sharmanthie Fernando \footnote{fernando@nku.edu}\\
{\small\it Department of Physics \& Geology}\\
{\small\it Northern Kentucky University}\\
{\small\it Highland Heights}\\
{\small\it Kentucky 41099}\\
{\small\it U.S.A.}\\

\end{center}

\begin{center}
{\bf Abstract}
\end{center}

In this paper, we have presented interesting properties of the static charged Born-Infeld-de Sitter black hole. They can have time-like as well as space-like singularities depending on the parameters of the theory.  The degenerate black holes lead to cold, ultra cold and Nariai black holes. The geometry of such black holes are discussed. A comparison is done with the Reissner-Nordstrom-de Sitter black holes.

\hspace{0.7cm}


\section{ Introduction}

Cosmological constant ($\Lambda$) was introduced by Einstein as a method to obtain a static universe from General Theory of Relativity. Later, when experiments revealed that the universe is  expanding, Einstein decided to  remove $\Lambda$ from the equations. In several observations in the recent history, it has been found that not only the universe is expanding, but  the expansion  is accelerating \cite{perl}\cite{reis}. The acceleration is driven by  some mysterious dark energy which is not  well understood yet. One of the proposals for the dark energy is the cosmological constant with a state parameter $ \omega = -1$ and negative pressure $ p = - \rho$. However, in order to agree with the observable data, the cosmological constant has to be fine tuned up to 120 orders of magnitude \cite{rapha}.\\

Born-Infeld theory of non-linear electrodynamics was proposed by Max Born and Leopold Infeld  \cite{born} to cure the divergences of the self energy of a charged point like particle in Maxwell's electrodynamics. Born-Infeld theory  has received interest in the recent past due to its relation to string theory \cite{leigh}\cite{gib1}\cite{tsey2}.
In Born-Infeld electrodynamics, the electric field of a point-like charge is given by $E = Q/ \sqrt{ r^4 + \frac{Q^2}{\beta^2}}$ which is clearly finite at $ r =0$. Hence its total energy is also finite e.g. \cite{rasheed2}. The maximum value of $E$ is $\beta$. From the string theory point of view, the maximum field strength $ \beta = \frac{ 1 } { 2 \pi \alpha'}$, where $\alpha'$ is the inverse of the string tension. Born-Infeld theory has exact $SO(2)$ electromagnetic duality inspite of  it's non-linear nature \cite{rasheed1}.

There are many interesting black hole solutions related  to Born-Infeld electrodynamics in the literature. Asymptotically flat charged black hole solutions in Born-Infeld gravity was presented by Breton in \cite{nora1}. Stability properties of the asymptotically flat charged black hole in Born-Infeld gravity were studied by the current author in \cite{fer1}\cite{fer2} \cite{fer3}. Melvin universe type solutions coupled to Born-Infeld electrodynamics were studied by Gibbons and Herderio \cite{gib2}. Exact solutions to Lovelock-Born-Infeld black holes were studied by Aiello et.al \cite{ello}.  Non-abelian black  hole solutions with Born-Infeld electrodynamics in higher dimensions were studied  in \cite{habib}.  Attractor mechanism for extreme black hole solutions in Einstein-Born-Infeld-dilaton gravity was studied by Gao \cite{gao}. Black hole solutions in Born-Infeld action coupled to Einstein gravity with a cosmological constant  has been found by several authors in \cite{cat}\cite{fer4}\cite{dey}\cite{cai}. In this paper, we focus on a static  Born-Infeld black hole with a positive cosmological constant.

The paper is organized as follows: In section 2, an introduction to the Born-Infeld-de Sitter black hole is given followed by a comparison with the Reissner-Nordstrom-de Sitter black hole in section 3. In section 4, the extreme black holes are discussed. In section 5, the topology is described for the degenerate horizons. Finally, the conclusion is given in section 6.

\section{ Born-Infeld-de Sitter black holes}

The Born-Infeld black hole is derived from the action,
\begin{equation} \label{action}
S = \int d^4x \sqrt{-g} \left[ \frac{R - 2 \Lambda)}{16 \pi G} + L(F) \right]
\end{equation}
where the  function  $L(F)$ is given by,
\begin{equation}
L(F) = 4 \beta^2 \left( 1 - \sqrt{ 1 + \frac{ F^{\mu \nu}F_{\mu \nu}}{ 2 \beta^2}} \right)
\end{equation}
Here, $\beta$ has dimensions $length^{-2}$ and $G$  $length^2$. In the paper, we will assume, $16 \pi G = 1$. Note that when $\beta \rightarrow \infty$, the Lagrangian $L(F) \rightarrow -F^2$ which corresponds to  Maxwell's  electrodynamics.
\noi
The static charged, spherical symmetric black hole derived from the action in eq.$\refb{action}$ is given by the metric,
\begin{equation} \label{metric}
ds^2 = -f(r) dt^2 + f(r)^{-1} dr^2 + r^2 ( d \theta^2 + sin^2\theta d \varphi^2)
\end{equation}
with,
\begin{equation}
f(r) = 1 - \frac{2M}{r} - \frac{  \Lambda r^2}{3} + \frac{2 \beta^2 r^2}{3} \left( 1 - \sqrt{ 1 + \frac{Q^2}{r^4 \beta^2}} \right) + \frac{ 4 Q^2}{ 3 r^2} \hspace{0.2cm}   _2F_1 \left( \frac{1}{4}, \frac{1}{2}, \frac{5}{4}, -\frac{Q^2}{ \beta^2 r^4} \right)
\end{equation}
Here $_2F_1$ is the hypergeometric function. The parameters in the metric are as follows; M is the mass, Q is the charge, $\beta$ is the non-linear parameter and $\Lambda$ is the cosmological constant. The electric field strength for the black hole is  given by,
\be
E = \frac{ Q}{ \sqrt{ r^4 + \frac{Q^2}{\beta^2} }}
\ee
Notice that $E$ is finite for $r \ra  0$ as expected.
\noi
In the limit $\beta \rightarrow \infty$, the elliptic integral can be expanded to give,
\begin{equation} \label{frn}
f(r)_{RN} = 1 - \frac{2 M}{r} + \frac{ Q^2}{r^2} - \frac{ \Lambda r^2}{3}
\end{equation}
which is the function  $f(r)$ for the Reissner-Nordstrom-de Sitter  black hole for Maxwell's electrodynamics.
Near the origin, the function $f(r)$ has the behavior,
\begin{equation}
f(r) \approx 1 - \frac{( 2M - A)}{r} - 2 \beta Q + \frac{ r^2}{3}  ( 2 \beta^2 - \Lambda)  + \frac{ \beta^3}{5}  r^4
\end{equation}
Here,
\begin{equation}
A = \frac{1}{3}\sqrt{ \frac{\beta}{ \pi} } Q^{3/2} \Gamma \left(\frac{1}{4} \right)^2
\end{equation}
Depending on  $M$ and $A$, it is possible to have three types of black hole solutions: \\

\noi
{\bf Case 1}: ($M > \frac{A}{2}$)\\
\noi
In this case, for $ r \ra 0$, $ f(r) \ra - \infty$. Hence the black hole will be similar to the Schwarzschild-de Sitter black hole. The function $f(r)$ is plotted in the Fig.1.\\

\noi
{\bf Case 2}: ($ M < \frac{A}{2}$)\\
\noi
In this case, for $ r \ra 0$, $ f(r) \ra  \infty$. Hence the black hole will be similar to the Reissner-Nordstrom-de Sitter black hole. The function $f(r)$ is plotted in the Fig.2.\\

\noi
{\bf Case 3}: ($ M = \frac{A}{2}$)\\
\noi
In this case, for $ r \ra 0$, $ f(r) \ra  ( 1- 2 Q \beta) $. Hence $f(r)$ is finite at $ r =0$ and single valued as given in Fig.3. The Kretschmann scalar still diverge at $ r =0$ for these black holes. So, the singularity exits. One could call them marginal black holes.

The above three cases can be described by a simple graph of $M$ vs $Q$ given in Fig.4.  Depending on the values of the parameters in the theory,   the function $f(r)$  can  have three roots, two roots, one root or none. When $f(r)$ has three roots, the behavior of $f(r)$  is similar to the Reissner-Nordstrom-de Sitter  black hole. When there are three roots, the smallest is the black hole inner horizon, the second largest is the black hole event horizon and the largest becomes the cosmological horizon. The graph in Fig.5 shows how for a given $M$, different number of horizons possible. When it has two roots, the black hole behave similar to the Schwarzschild-de Sitter  black hole. In this case, one will be the black hole horizon and the other will be the cosmological horizon. There are also black holes with degenerate horizons which will be discussed in detail later. Hence the Born-Infeld-de Sitter  black  hole is  interesting  since it possess  the characteristics of the most well known black holes in the literature.

\begin{center}
\scalebox{.9}{\includegraphics{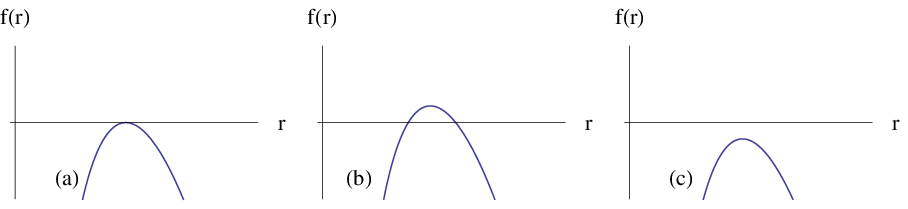}}

\vspace{0.3cm}
\end{center}

Figure 1. The figure shows the graphs for $f(r)$ vs $r$ for various masses $M$.  In this case, $M > A/2$. Here $Q= 0.2307$ ,  $\beta = 1.4$ and $ \Lambda = 0.5$. \\

\begin{center}
\scalebox{.9}{\includegraphics{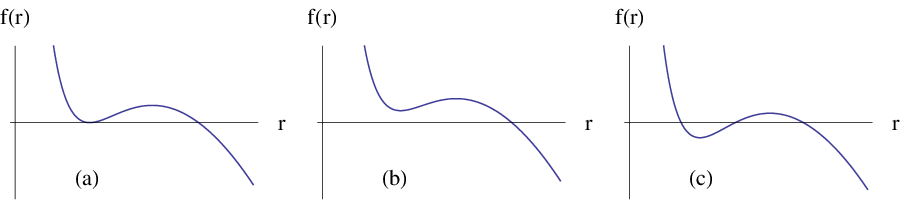}}

\vspace{0.3cm}

\scalebox{.9}{\includegraphics{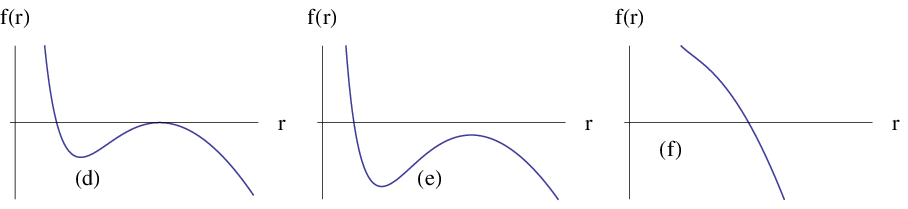}}
\end{center}

Figure 2. The figure shows the graphs for $f(r)$ vs $r$ for vrious values of $M$ and $Q$.  Here $ M < A/2$ and    $\beta = 1.4$ and $ \Lambda = 0.5$. \\

\begin{center}
\scalebox{.9}{\includegraphics{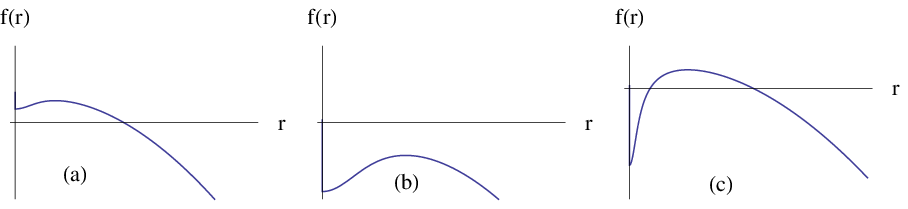}}

\vspace{0.3cm}
\end{center}

Figure 3. The figure shows the graphs for $f(r)$ vs $r$.  In all the cases, $M = A/2$. Here $\beta = 1.4$. $M, Q$ and $\Lambda$ are varied.

\begin{center}

\scalebox{.9}{\includegraphics{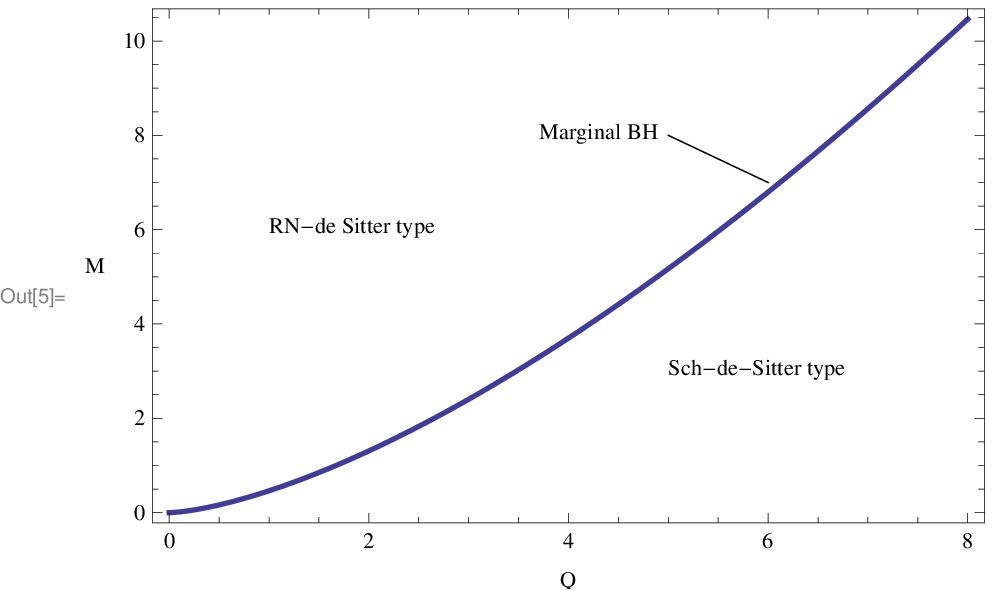}}

\end{center}

Figure 4. The figure shows the graphs for $M$ vs $Q$. Here, $Q = 0.5, M = 0.2$ and $\Lambda = 0.5$.\\

\begin{center}

\scalebox{.9}{\includegraphics{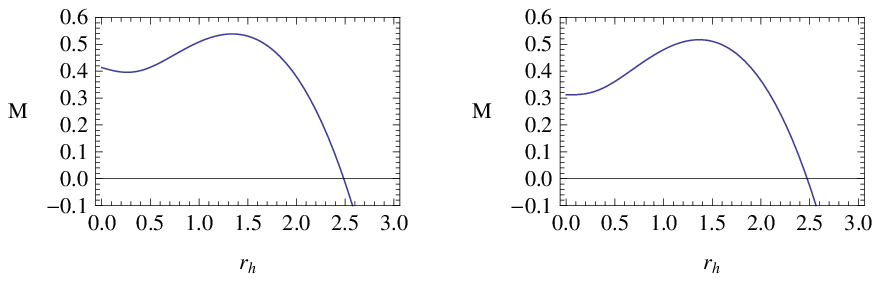}}

\end{center}

Figure 5. The figure shows the graphs for $M$ vs $r_h$ for $\Lambda =0.5, \beta = 1.4$. For the first graph $Q = 0.431$ and for the second graph $Q = 0.357$.\\

In Fig. 6, the black horizon radius is plotted against the non-linear parameter $\beta$. One can observe that for large $\beta$ the black hole is smaller.

\begin{center}

\scalebox{.9}{\includegraphics{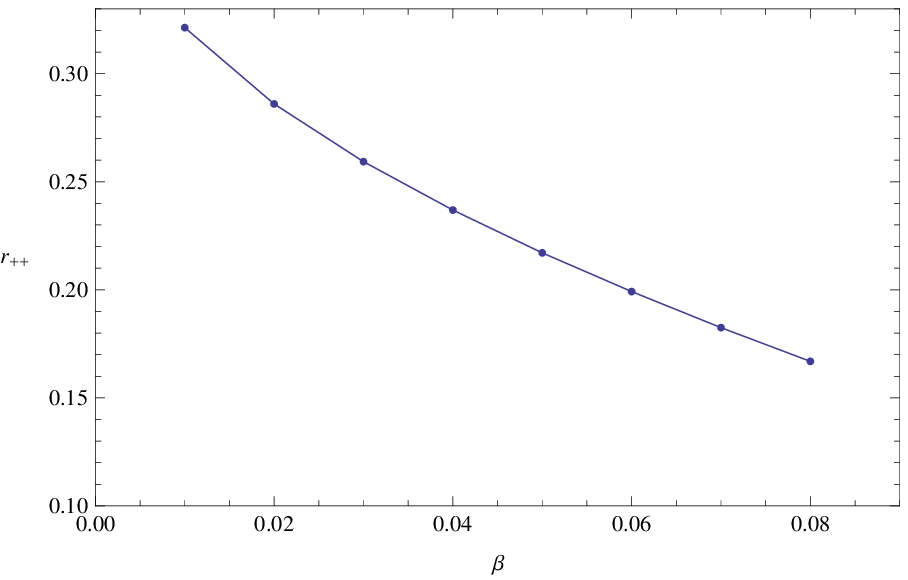}}

\end{center}

Figure 6. The figure shows the graphs for $r_{++}$ vs $\beta$. Here, $Q = 0.5, M = 0.2$ and $\Lambda = 0.5$.\\

The Hawking temperature for the Born-Infeld black hole, $T$ is obtained by the usual relation, $ T = \frac{f'(r)}{ 4 \pi}$ as,
\begin{equation}
T = \frac{1}{4\pi r_h} \left[  1  + (2\beta^2 - \Lambda)r_h^2 -  2 \beta {\sqrt{(Q^2 + r_h^4 \beta^2)}}  \right]
\end{equation}
In the Fig.7, the temperature for the cosmological horizon and the black hole event horizon  is plotted vs $\beta$. The black hole considered in this case is Schwarzschild-de Sitter type. From the graphs, the temperature for the both horizons  increase with $\beta$.

\begin{center}

\scalebox{.9}{\includegraphics{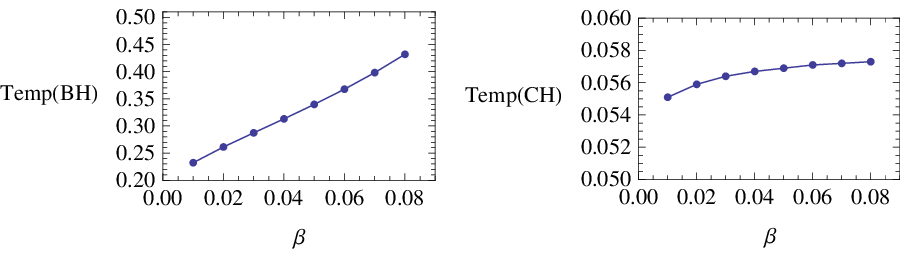}}

\end{center}

Figure 7. The figure shows the graphs for $Temp$ vs $\beta$. The first graph shows the temperature for the black hole horizon while the second graph shows the temperature for the cosmological horizon. Here, $ Q = 0.5, M = 0.2$ and $ \Lambda = 0.5$.\\


\section{Comparison of the Born-Infeld-de Sitter black hole with the Reisnner-Nordstrom-de Sitter black hole}

In order to fully appreciate the properties of the Born-Infeld-de Sitter black hole, it is important to compare it with the charged black hole in the Maxwell's electrodynamics, which is Reissner-Nordstrom-de Sitter black hole whose geometry is defined by the metric in eq.$\refb{metric}$ with $f(r)$ given by eq.$\refb{frn}$. In the Fig.8, the functions $f(r)$ for both black holes are plotted. It is observed that while the cosmological horizon are close ( not the same), the inner black hole horizon is smaller for the Born-Infeld-de Sitter black hole.

\begin{center}

\scalebox{.9}{\includegraphics{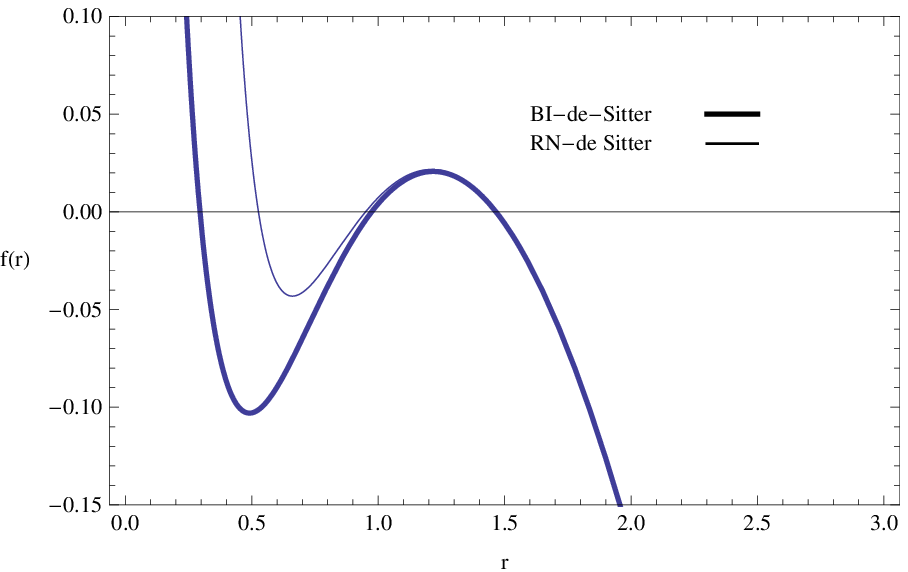}}

\end{center}

Figure 8. The figure shows the graphs for $f(r)$ vs $r$ for the Born-Infeld-de Sitter and the Reissner-Nordstrom-de Sitter black hole for the same mass, charge and the cosmological constant.   Here, $\beta = 1.4, Q = 0.6, \Lambda = 0.5$ and $ M = 0.593$\\

We would like to review some interesting properties of the Reissner-Nordstrom-de Sitter black holes here. Such black holes could have three horizons, $ r_+, r_{++}$ and $r_c$. The first two are black hole inner and outer horizons while the third is the cosmological horizon.  In general, the black hole outer horizon and the cosmological horizon are not in thermal equilbrium. The only time this happens is when there is a Nariai type degeneracy (as given in Fig. 2(d)) or in the lukewarm black holes \cite{jerzy}.
A description of thermodynamics and instantons of Reissner-Nordstrom-de Sitter black holes are given in \cite{mann2}\cite{piazzo}\cite{romans} \cite{ross} \cite{mann}.  Nariai black holes in other theories are discussed in \cite{diaz} \cite{raphael3}. Nariai black holes in higher dimensions are discussed  by Cardoso et.al. in \cite{lemos}.

The Hawking temperature is plotted for  both black holes in Fig.9.  From the Figure, it is clear that the Born-Infeld-de Sitter black hole is hotter than its counterpart in Maxwell's electrodynamics.

\begin{center}

\scalebox{.9}{\includegraphics{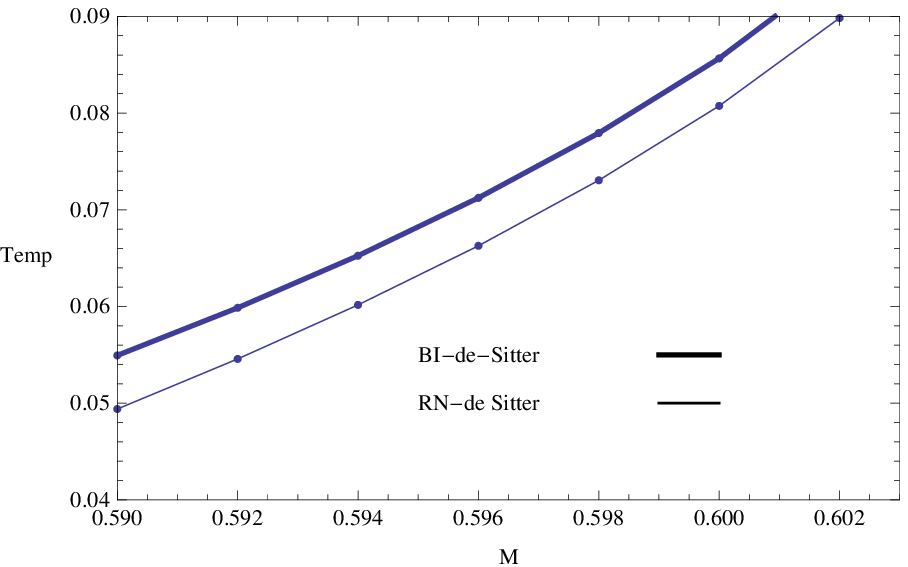}}

\end{center}

Figure 9. The figure shows the graphs for $Temp$ vs $M$ for the Born-Infeld-de Sitter and the Reissner-Nordstrom-de Sitter black hole for the same charge and the cosmological constant.   Here, $\beta = 1.4, Q = 0.6$ and $\Lambda = 0.5$.\\


\section{ Extreme black holes}

The main focus in this section is to study the degenerate horizons of the Born-Infeld-de Sitter black holes. For degenerate horizons, $ f(r) = f'(r)=0$. By solving these two equations one arrives at,
\be
r_{ex}^2 = \frac{ (2 \beta^2 - \Lambda) \pm \sqrt{\beta^4 - 4 Q^2 \beta^4 \Lambda + Q^2 \beta^2 \Lambda} }
{ \Lambda (  4 \beta^2 - \Lambda)}
\ee
We can also write $r_{ex}^2$ as,

\be
r_{ex}^2 = \frac{  (2 \beta^2 -  \Lambda) \pm \sqrt{\delta} }{ \Lambda ( 4 \beta^2 - \Lambda)}
\ee
where
\be \label{delta}
\delta = ( 2 \beta^2 - \Lambda)^2 + \Lambda ( 1 - 4 \beta^2 Q^2) ( 4 \beta^2 - \Lambda)
\ee
It may look  $r_{ex}$ to be  independent of the mass. However, the mass of the corresponding black hole will be obtained by using $f(r_{ex})=0$.

Now, we will study the type of horizons for various values of the parameters in the theory. 

\subsection{ ($2 \beta^2 - \Lambda) >0$}

In this case, $ (4 \beta^2 - \Lambda) >0$, Hence $ \delta >0$ and $\sqrt{\delta}$ is real. \\

\noi
{\bf Case1: $ (1 - 4 \beta^2 Q^2) >0$ or $  Q \beta < \frac{1}{2}$}\\

\noi
In this case $ \sqrt{\delta} > ( 2 \beta^2 - \Lambda)$. Hence $r_{ex}^2$ will have only one root with the $``+''$ sign in front of the $\sqrt{\delta}$.  Hence, in this case, the black hole will be degenerate Schwarzschild-de Sitter type as given in the Fig.10.

\begin{center}

\scalebox{.9}{\includegraphics{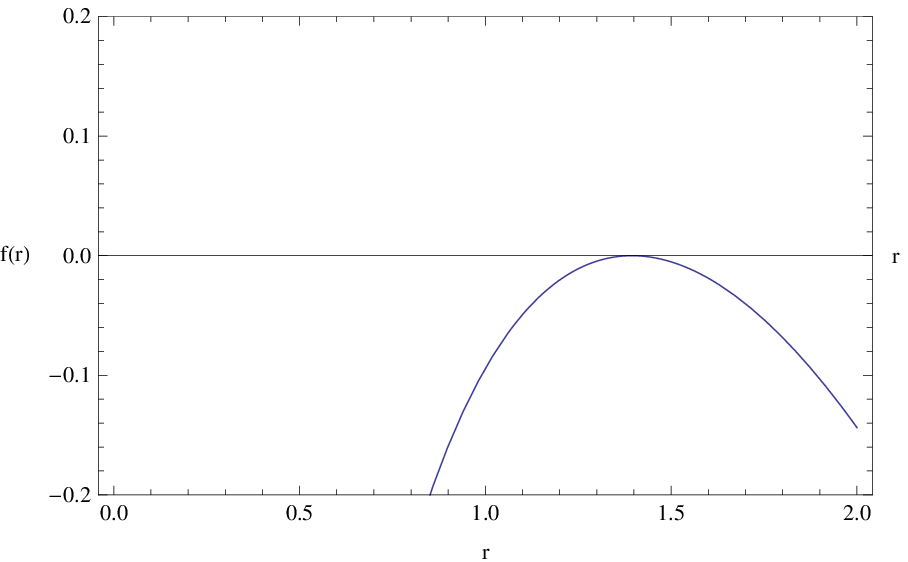}}

\end{center}

Figure 10. The figure shows the graphs for $f(r)$ vs $r$ for a degenerate black hole of the Schwarzschild-de Sitter type. Here, $\beta = 1.4, Q = 0.2307, \Lambda = 0.5$ and $ M = 0.49036$.\\

{\bf Case 2:  $ (1 - 4 \beta^2 Q^2) <0$, or $  Q \beta > \frac{1}{2}$}\\

\noi
In this case, $ \sqrt{\delta} < ( 2 \beta^2 - \Lambda)$. Hence there will be two values for $r_{ex}^2$ with the $+$ and the $-$ in front of $\sqrt{\delta}$. Hence, in this case, the black hole will be degenerate Reissner Nordstrom-de Sitter type as given in the Fig.11.

\begin{center}

\scalebox{.9}{\includegraphics{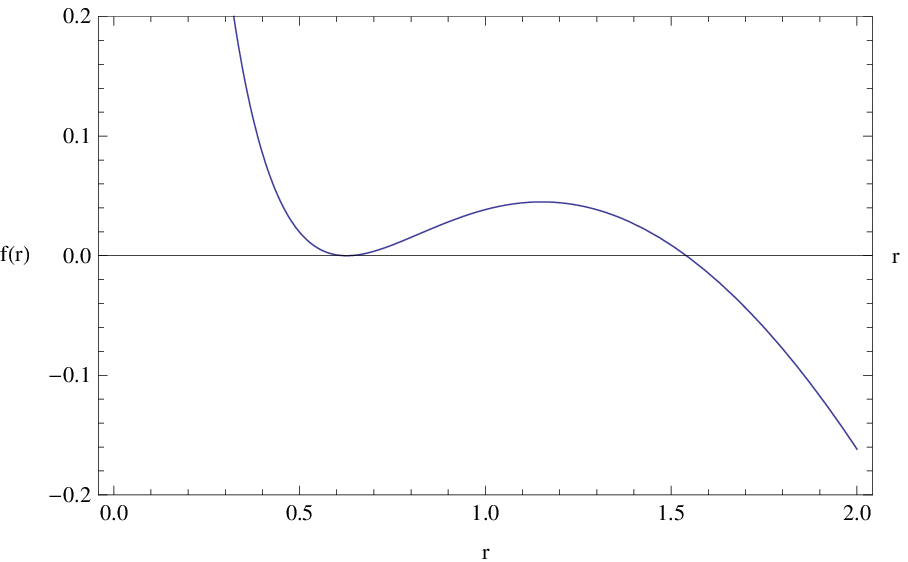}}

\scalebox{.9}{\includegraphics{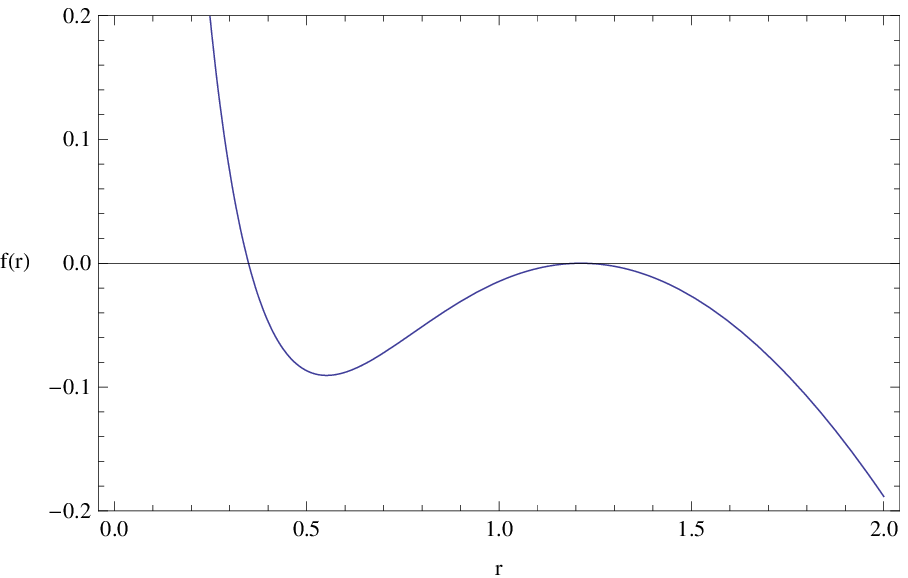}}

\end{center}

Figure 11. The figure shows the graphs for $f(r)$ vs $r$ for a degenerate black hole of the Reissner-Nordstrom-de Sitter type. Here, $\beta = 1.4, Q = 0.2307$ and $ \Lambda = 0.5$.

\noi
{\bf Case 3: $ (1 - 4 \beta^2 Q^2) =0$, or $ Q \beta = \frac{1}{2} $}\\

\noi
In this case, $ \sqrt{\delta} = ( 2 \beta^2 - \Lambda)$. Hence,
\be
r_{ex}^2 = 0; ~~ r_{ex}^2 = \frac{ 2 ( 2 \beta^2 - \Lambda)}{ \Lambda ( 4 \beta^2 - \Lambda)}
\ee
The black hole in this case is degenerate Schwarzschild-de Sitter type as given in the Fig.10.

\subsection{$ Q = \frac{\beta}{ \sqrt{ \Lambda ( 4 \beta^2 - \Lambda)} } $}

In this case $ \delta =0$. Hence,
\be
r_{ex}^2  = \frac{  ( 2 \beta^2 - \Lambda)}{ \Lambda ( 4 \beta^2 - \Lambda)}
\ee
This is indeed a triple root of $f(r)$ since $f''(r)=0$ for this case as shown in Fig.12.

\begin{center}

\scalebox{.9}{\includegraphics{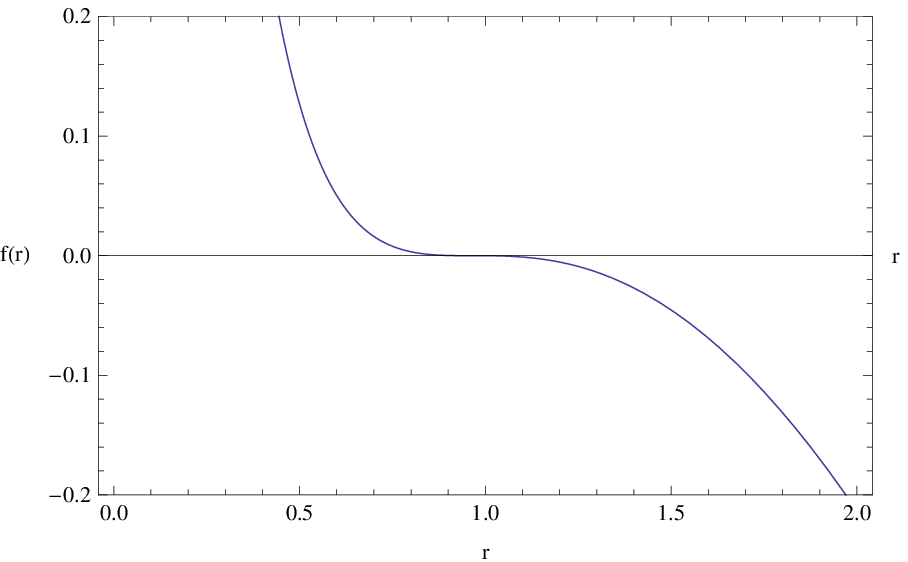}}

\end{center}

Figure 12. The figure shows the graphs for $f(r)$ vs $r$ for a degenerate black hole of the Reissner-Nordstrom-de Sitter type. Here, $\beta = 1.4, Q = 0.7307, \Lambda = 0.5$ and $ M = 0.6803$.\\

\subsection{($4 \beta^2 > \Lambda > 2 \beta^2$)}

Here, $(2 \beta^2 - \Lambda) < 0$ and $ (4 \beta^2 - \Lambda) > 0$. There are two cases to consider:\\

\noindent
{\bf  Case 1: $Q \beta < \frac{1}{2}$}\\

In this case, by observing the expression for $\delta$ in eq.$\refb{delta}$, it is clear that $\sqrt{\delta} > | ( 2 \beta^2 - \Lambda)|$. Hence only $+$ will give a real value for $r_{ex}$ as,
\be
r_{ex}^2 = \frac{ 2 \beta^2 - \Lambda +  \sqrt{\delta} }{ \Lambda ( 4 \beta^2 - \Lambda)}
\ee
The extreme black hole will be Schwarzschild-de Sitter type as given in Fig. 10.\\

\noindent
{\bf Case 2: \bf $Q \beta > \frac{1}{2}$}\\

In this case for $\delta$ to be positive, $Q < \frac{ \beta^2}{ \sqrt{ 4 \beta^4 \Lambda - \beta^2 \Lambda^2}}$. If both conditions are satisfied, $\sqrt{\delta} < | ( 2 \beta^2 - \Lambda)|$. Since $ ( 2 \beta^2 - \Lambda)<0$, there wont be a real solution for $r_{ex}$. Hence no extreme solution for such parameters.

\subsection{ ( $\Lambda > 4 \beta^2$)}

Here, $ (2 \beta^2 - \Lambda) < 0$ and $ (4 \beta^2 - \Lambda) < 0$. By looking at the eq.$\refb{delta}$, one can see that $\delta$ is always positive. Again there are two cases to consider:\\

\noindent
{\bf Case1 : \bf $Q \beta > \frac{1}{2}$}\\

\noindent
In this case, $ \sqrt{\delta} > |( 2 \beta^2 - \Lambda)|$. For $r_{ex}$ to be real, the $-$ sign has to be chosen in front of $\sqrt{\delta}$. Hence it seems that there is one root for $r_{ex}$ leading to Schwarzschild-de Sitter type extreme black hole with,
\be
r_{ex}^2 = \frac{  (2 \beta^2 -  \Lambda) -  \sqrt{\delta} }{ \Lambda (  4 \beta^2 - \Lambda)}
\ee
However, when the mass is calculated for the parameters, it was noticed that $ M < A/2$. Hence the solution has to be Reissner-Nordstrom-de Sitter type according to the discussion in section(2). In fact when we plotted the graph there were no extreme black holes; it was a Reissner-Nordstrom-de Sitter type black hole with a single horizon which was not a degenerate one. Hence the conclusion is that for the above conditions, there are no extreme black holes.\\

\noindent
{\bf Case 2: \bf $Q \beta < \frac{1}{2}$}\\

\noindent
In this case, $ \sqrt{\delta} < |( 2 \beta^2 - \Lambda)|$. For $r_{ex}$ to be real, both signs can  be chosen in front of $\sqrt{\delta}$. Hence it seems that there are two roots for $r_{ex}$ leading to Reisnner-Nordstrom-de Sitter type extreme black hole. However, when the mass is calculated, $ M > A/2$ for one of the roots which  signals that the solutions should be Schwarzschild-de Sitter type. For the other root, $ M < A/2$ which signals the solutions to be Reissner-Nordstrom-de Sitter type. The solution  which works  is the first one given by,
\be
r_{ex}^2 = \frac{  (2 \beta^2 -  \Lambda) +  \sqrt{\delta} }{ \Lambda (  4 \beta^2 - \Lambda)}
\ee
leading to Schwarzschild-de Sitter type extreme black hole.\\

\noindent
{\bf  Case 3: $Q \beta = \frac{1}{2}$}\\

\noindent
In this case, $ \sqrt{\delta} = |( 2 \beta^2 - \Lambda)|$. Hence there is one non-zero real value for $r_{ex}$ as,
\be
r_{ex}^2 = \frac{ 2 (2 \beta^2 -  \Lambda)}{ \Lambda (  4 \beta^2 - \Lambda)}
\ee
This is a Schwarzschild-de Sitter type extreme black hole.


\section{Topology of the Extreme black holes}

\subsection{ Nariai type black holes}

In  Nariai black holes, the cosmological horizon and the black hole event horizon coincide. For a nearly extreme black hole of this type, one can approximate the function $f(r)$ as \cite{paw},
\begin{equation}
f(r) = \frac{ f''(r_{ex})}{2} ( r - r_1) ( r - r_2)
\end{equation}
Here, $r_1$ and $r_2$ represents the two close horizons, $r_{++}, r_c$. One can introduce new coordinate $ \chi$ as,
\begin{equation}
r = r_{ex} + \epsilon cos \chi
\end{equation}
around the close horizons. Here $\epsilon$ is small. Hence, $\chi=0$ corresponds to $r_1$ and $\chi = \pi$ corresponds to $r_2$. A new time-like coordinate $\psi$ is also introduced such as,
\begin{equation}
t = \frac{ 2 \psi}{ \epsilon f''(r_{ex}) }
\end{equation}
Note that $ f''(r_{ex}) < 0$ for this type of black holes due to the nature of the function $f(r)$ at $ r = r_{ex}$. Now, by substituting the new coordinates, the metric can be written as,
\be
ds^2 = \frac{ -2}{ f''(r_{ex})} \left( - sin^2\chi d \psi^2 + d \chi^2 \right) + r_{ex}^2 d \Omega^2
\end{equation}
The above geometry corresponds to $ dS_2 \times S^2$. The $dS_2$ has a scalar curvature,  
\be
 R_{dS_2}=    |f''(r_{ex})|
 \ee
For the Born-Infeld black hole,
\be \label{fdouble}
f''(r_{ex}) = \frac{ Q^2 ( 4 \beta^2 - \Lambda) - 2 r_{ex}^4 \beta^2 \left( \Lambda - 2 \beta^2 + 2 \beta^2 \sqrt{ 1 + \frac{ Q^2}{ r_{ex}^4 \beta^2}} \right) } { Q^2 + r_{ex}^4 \beta^2}
\ee
Reissner-Nordstrom-de Sitter black hole also has the same topology near the degenerate horizon.

Even though the horizons coincide in the Schwarzschild-like coordinates, the proper distance between them is not-zero. it can be calculated in the new coordinate system as,
\be
\int^{\pi}_{0} \frac{ \sqrt{2} d \chi}{ \sqrt{ |f''(ex)|} }= \frac{ \sqrt{2}\pi}{ \sqrt{|f''(ex)|}}
\ee
Also, the Hawking temperature, which is defined by $ T = \frac{\kappa}{ 2 \pi}$ with $\kappa$ being the surface gravity, seems like zero with the usual definition $\kappa = \frac{ f'(r_{ex})}{2}$. However, Cho and Nam \cite{nam2} redefined the surface gravity for the Nariai-type black holes as,
\be
\tilde{\kappa} = \frac{ f''(r_{ex})}{2}
\ee
Hence the temperature at the degenerate horizon is,
\be
\tilde{T} = \frac{\sqrt{f''(r_{ex})}}{ 2 \sqrt{2} \pi}
\ee
with $f''(r_{ex})$ given by eq.\refb{fdouble}.

\subsection{ Cold black holes}

In cold black holes, the black hole inner horizon and the black hole outer horizon coincides. This is clearly  given in the figure (a) in Fig.2. In cold black holes $ f(r) = f')r)=0$ at the degenerate horizon. To understand the topology of the cold black holes, we can choose a new coordinate $y$ to describe the near extreme geometry as \cite{nam},
\be
r = r_{ex} - \epsilon y
\end{equation}
The function $f(r)$ can be expanded around $r = r_{ex}$ as,
\begin{equation}
f(r) \approx  \frac{f''(r_{ex})}{2} ( \epsilon y)^2
\end{equation}
A  new time coordinate is  defined as $ \psi = \epsilon t $.  With these new coordinates, the metric is approximated to be,
\begin{equation}
ds^2 = \frac{ -f''(r_{ex})}{2} y^2 d \psi ^2 + \frac{ 2}{ f''(r_{ex})} \frac{ dy^2}{y^2}  + r_{ex}^2  d \Omega^2
\end{equation}
Since $ f''(r_{ex}) >0$ for the cold black hole, the above geometry represents $AdS_2 \times S^2$ topology. The $AdS_2$ has the curvature $-f''(r_{ex})/2$. The topology is the same for the cold Reissner-Nordstrom-de Sitter black hole but with curvature  $\Lambda$.\\

\subsection{ Ultra-cold black holes}
From section(4), for the  ultra cold black hole, all three horizons coincide as,
\be
r_{+} = r_{++} = r_{c} = \sqrt{ \frac{ ( 2 \beta^2 - \Lambda)}{ \Lambda ( 4 \beta^2 - \Lambda)}}
\ee
Here, $ f(r_{ex} )= f'(r_{ex}) = f''(r_{ex})=0$.\\

To understand the geometry near the horizon, a new coordinate is defined for the near extreme case as,  $ y = \eta \sqrt{ f''(r_{ex})/2}$. One can substitute the new coordinate and take the limit $f''(r_{ex}) \rightarrow 0$ which leads to,
\begin{equation}
ds^2 =  - \eta^2 d \psi^2 +  d \eta^2 + r_{ex}^2 d \Omega^2
\end{equation}
The above geometry has the topology, $ R^2 \times S^2$. This is similar to the topology of the  ultra-cold Reissner-Nordstrom-de Sitter black hole near the degenerate horizon. The temperature is zero.


\section{Conclusions}

In this paper, we have studied the properties of the Born-Infeld-de Sitter black holes. Due to the presence of the cosmological constant, the black holes have  a cosmological horizon. The black holes in Born-Infeld-de Sitter gravity are interesting in the sense that there are variety of possibilities when it comes to the number of horizons and the nature of the singularity. Depending on the mass, charge, and the non-linear parameter, the black holes could have up-to three horizons. These could be similar to Reissner-Nordstrom-de Sitter, Schwarzschild-de Sitter or  a marginal black hole with finite $f(r)$. It is also possible to have naked singularities. The non-linear nature of the electrodynamics change the singularity at the origin drastically. The behavior of the function $f(r)$ is dominated by $M/r$ term rather than a $Q^2/r^2$ term as in Maxwell's electrodynamics couple to gravity. In the Reissner-Nordstrom-de Sitter black hole, the singularity is time-like, while in the Born-Infeld counterpart, it could be time-like or  space-like depending on the parameters of the theory. Also the Born-Infeld black holes are colder.

In this paper, we have focused on the extreme black hole with degenerate horizons. A thorough analysis is done as to what type of degenerate horizon would emerge for the parameters of the theory. There are cold black holes where the black hole inner horizon and the outer horizon merge with zero temperature. These black holes have topology $AdS_2 \times S^2$.  There are Nariai  type black holes where the black hole outer horizon and the cosmological horizon merge with the topology, $dS_2 \times S^2$. The temperature and the distance between the horizons are calculated for such black holes. The ultra cold black holes has the topology, $ R^2 \times S^2$. The geometries of these extreme black holes are similar to the  corresponding ones 
in the Reissner-Nordstrom-de Sitter black holes. These topologies are discussed in mored detail in the book by Griffths $\&$ J. Podolsk$\acute{y}$ \cite{pod}.

As for future work, it  would be interesting to study the thermodynamics of these interesting class of black holes. In a recent paper, Dolan et.al. \cite{dolan} did an analysis of de Sitter black holes considering the cosmological constant as a thermodynamical variable. It would be interesting to use  such an approach to study thermodynamical properties of Born-Infeld-de Sitter black holes. It is also interesting to study the possibility of pair creation similar to what happens in Reissner-Nordstrom-de Sitter black holes \cite{mann}.  We have not discussed lukewarm black holes where the black hole horizon and the cosmological horizon has the same temperature \cite{jerzy}. That would be an interesting aspect to study.



\begin{thebibliography}{99}





\bibitem{perl} S. Perlmutter et. al., {\it Measurements of $\Omega$ and $\Lambda$ from 42 high-redshift supernovae}, Astrophys. J. {\bf 517} 565 (1999)

\bibitem{reis} A.G. Riess  et. al., {\it Observational evidence from Supernovae for an accelerating universe and a cosmological constant},  Astron. J. {\bf 116} 1009 (1998);
{\it BVRI Light curves for 22 Type Ia Supernovae}, Astron. J. {\bf 117} 707(1999)


\bibitem{rapha} R. Bousso, {\it TASI lectures on the cosmological constant}, arXiv:0708.4231


\bibitem{born} M. Born and L. Infeld,  {\it Foundations of the new field theory},  Proc. Roy. Soc. Lond. {\bf A144} (1934) 425.


\bibitem{leigh} R. G. Leigh,  {\it Dirac-Born-Infeld action from Dirchlet $\sigma$-model}, Mod. Phys. Lett {\bf A4} (1989) 2767.

\bibitem{gib1} G. W. Gibbons, {\it Aspects of Born-Infeld Theory and String/M-Theory},  Rev.Mex.Fis.49S1:19-29 (2003),  hep-th/0106059

\bibitem{tsey2} A. A. Tseytlin, {\it Vector field effective action in the open superstring theory}, hep-th/9908105

\bibitem{rasheed2} D. A. Rasheed, {\it Non-linear  electrodynamics: zeroth and first laws of black hole mechanics}, hep-th/9702087


\bibitem{rasheed1} G. W. Gibbons \& D. A. Rasheed, {\it Electromagnetic duality rotations in non-linear electrodynamics}, hep-th/9506035


\bibitem{nora1} N. Breton, {\it Born-Infeld generalizations of the Reissner-Nordstrom black hole}, gr-qc/0109022

\bibitem{fer1} S. Fernando, {\it Gravitational perturbation and quasi-normal modes of charged black holes in Einstein-Born-Infeld gravity}, Gen. Rela. Grav. {\bf 37} 585(2005)

\bibitem{fer2} S. Fernando \& C. Holbrook, {\it Stability and quasi normal modes of charged black holes in Born-Infeld gravity},  Int. Jour. Theo. Phys. {\bf 45} 1630 (2006)

\bibitem{fer3} S. Fernando,  {\it Decay of massless Dirac field around the Born-Infeld black hole}, Int. Jour. Mod. Phys.{\bf A 25} (2010) 669

\bibitem{gib2} G. W. Gibbons and C. A. R. Herdeiro, {\it The Melvin Universe in Born-Infeld theory and other theories of non-linear electrodynamics}, Class. Quant. Grav. {\bf 18} (2001) 1677

\bibitem{ello} M. Aiello, R. Ferro \& G. Girbet, {\it Exact solutions of Lovelock-Born-Infeld black holes}, Phys. Rev. {\bf D 70} 104014 ( 2004)

\bibitem{habib} S. H. Mazharimousavi, M. Halisoy \& Z. Amirabi, {\it New non-abelian black hole solutions in Born-Infeld gravity}, Phys. Rev. {\bf D78} (2008) 064050

\bibitem{gao} X. Gao, {\it Non-supersymmetric attractors in Born-Infeld black holes with a cosmological constant}  JHEP, 0711:006 ( 2007)



\bibitem{cat}
    M. Cataldo and A. Garcia,  {\it Three dimensional black hole coupled to
the Born-Infeld electrodynamics},  Phys. Lett. B {\bf 456}, 28
(1999) 

\bibitem{fer4}
    S. Fernando and D. Krug,  {\it Charged black hole solutions in
Einstein-Born-Infeld gravity with a cosmological constant},  Gen.
Relativ. Gravit. {\bf 35}, 129 (2003)

\bibitem{dey}
    T. K. Dey, {\it Born-Infeld black holes in the presence of a
cosmological constant},  Phys.Lett. B {\bf 595}, 484 (2004).


\bibitem{cai}
    Rong-Gen Cai, Da-Wei Pang and Anzhong Wang, {\it Born-Infeld black
holes in (A)dS spaces}, Phys. Rev.D {\bf 70}, 124034 (2004)



\bibitem{jerzy} J. Matyjasek \& K. Zwierzchowska, {\it Lukewarm black holes in quadratic gravity} Mod. Phys. Lett. {\bf A26} 999 (2011)


\bibitem{mann2} D. Astefanesei, R.B. Mann \& E. Radu, {\it Reissner-Nordstrom-de Sitter black hole, planar coordinates and dS/CFT}, JHEP {\bf 029} 0401 (2004)

\bibitem{piazzo} F. Belgiorno, S.L. Caciatori \& F.D. Piazza, {\it Pair-production of charged Dirac particles on charged Nariai and ultracold black hole manifolds}, JHEP {\bf 028} 0908 (2009)


\bibitem{romans} L.J. Romans, {\it Supersymmetric, cold and lukewarm black holes in cosmological Einstein-Maxwell theory}, Nucl. Phys. {\bf B383} 395 (1992)


\bibitem{ross} S.W. Hawking \& S.F. Ross, {\it Duality between electric and magnetic black holes}, Phys. Rev. {\bf D52} 5865 (1995)

\bibitem{mann} R.B. Mann \& S.F. Ross, {\it Cosmological production of charged black hole pairs}, Phys.Rev. {\bf D52} 2254 (1995)



\bibitem{diaz} P. Diaz \& A Segui, {\it Generalized Nariai solutions for Yang-type monopoles}, Phys. Rev. {\bf D76} 064033 (2007)


\bibitem{raphael3} R. Bousso, {\it Charged Nariai black hole with a dilaton}, Phys. Rev. {\bf D55} 3614 (1999)

\bibitem{lemos} V. Cardoso, O. J. C. Dias, \& J. P. S. Lemos, {\it Nariai, Bertotti-Robinson and anti-Nariai solutions in higher dimensions}, Phys. Rev. {\bf D 70} 024002 (2004) 

\bibitem{paw} J. Matyjasek, P. Sadurski \& D. Tryniecki, {\it Inside the degenerate horizons of the regular black holes},  arXiv:1304.6347
\bibitem{nam2} J. Cho \& S. Nam, {\it The entropy function for the black holes of Nariai class}, JHEP 0803:027, (2008)


\bibitem{nam} J. Cho \& S. Nam, {\it Non-supersymmetric attractor with the Cosmological constant}, JHEP 0707:011, (2007)




\bibitem{pod} J. B. Griffths \& J. Podolsk$\acute{y}$, {\it Exact space-times in Einstein's General Relativity}, Cambridge Monographs on Mathematical Physics, ( 2009).

\bibitem{dolan} B. P. Dolan, D. Kastor, R. B. Mann \& J. Traschen, arXiv:1301.5926







\end{thebibliography}
\end{document}